# Congruence and Plausibility, not Presence?! Pivotal Conditions for XR Experiences and Effects, a Novel Approach

**Marc Erich Latoschik[1*], Carolin Wienrich[2]**

[1]Human-Computer Interaction Group, University of Würzburg, Würzburg, Germany

[2]Human-Technology-Systems Group, University of Würzburg, Würzburg, Germany

**\* Correspondence:**
Marc Erich Latoschik
marc.latoschik@uni.wuerzburg.de

**Keywords: XR, experience, presence, congruence, plausibility, coherence, theory, prediction**

**Abstract**

Presence often is considered the most important quale describing the subjective feeling of being in a computer-generated and/or computer-mediated virtual environment. The identification and separation of orthogonal presence components, i.e., the place illusion and the plausibility illusion, has been an accepted theoretical model describing Virtual Reality (VR) experiences for some time. This perspective article challenges this presence-oriented VR theory. First, we argue that a place illusion cannot be the major construct to describe the much wider scope of Virtual, Augmented, and Mixed Reality (VR, AR, MR: or XR for short). Second, we argue that there is no plausibility illusion but merely plausibility, and we derive the place illusion caused by congruent and plausible generation of spatial cues, and similarly for all the current model's so-defined illusions. Finally, we propose congruence and plausibility to become the central essential conditions in a novel theoretical model describing XR experiences and effects.

## Introduction

"*A review and categorization of definitions of presence has demonstrated that it is an unusually rich and diverse concept. [...] Presence, and definitions of presence, touch on profound issues involving the nature of reality and existence; human cognition, affect and perception; the characteristics, uses and impacts of primitive, advanced and futuristic technologies; and the subtleties of interpersonal communication and human-technology interaction*" (Lombard and Jones, 2015, 30).

Lombard and Jones highlight the significance of the presence construct. However, they also reflect on the wide scope, the potential diversity of definitions, and hence the blurred concreteness of its very nature. There are other considerable problems with the presence construct. Biocca's book problem addresses the technology-driven interpretation since presence can be experienced by imagination and/or in narratives presented in non-immersive media like books (Schubert and Crusius, 2002). Then, presence models often expose a sole dependency on other qualia and constructs like the place, plausibility, and social presence illusions (Skarbez et al., 2017), or the virtual body ownership illusion (Latoschik et al., 2017; Waltemate et al., 2018). Even more, a central focus on a sense of "*being there*" for XR applications does not capture the essence of the many variations of XR covered by the Virtuality Continuum (Milgram and Kishino, 1994). In essence, if we want to guide designers and developers to create compelling XR applications and experiences as initially motivated by (Heeter, 1992), we need well-defined qualities to strive for, with pragmatic ways to operationalize modifications to these qualities, and to provide clear-cut entry-points for a user-centered design process.

**Related Work**

There now is a considerable body of knowledge on presence, see excellent overviews in (Lombard et al., 2015; Skarbez et al., 2017). We follow (Lombard and Jones, 2015) and start by defining **presence: The related quale mediated by XR-technology, i.e., the degree one believes that she exists within a mediated space** (Jerome and Jordan, 2007), including concepts of virtual presence (Heeter, 1992) and telepresence: "*The biggest challenge to developing telepresence is achieving that sense of 'being there.'*" (Minsky, 1980). Heeter concluded, „*A question to guide designers of virtual worlds is how do I convince participants that they and the world exist?*" (Heeter, 1992).

Slater and Wilbur proposed immersion as an objectively measurable (system) characteristics and stated that presence would be "*the potential psychological and behavioral response to immersion*" (Slater and Wilbur, 1997), opening up a pathway to (technically) manipulate presence experiences. Slater later proposed two orthogonal components of presence, the place illusion (PI) and the plausibility illusion (Psi) (Slater, 2009), a separation that received wide acceptance. Lately, Skarbez et al. extended on this model as depicted in Figure 1 (Skarbez et al., 2017). They define presence as the "*the perceived realness of a mediated or virtual experience*". They further integrate additional constructs into their model, namely copresence and social presence, and specify Psi as well as copresence also to affect social presence. Finally, regarding the level of objectively measurable characteristics affecting the different presence components, they claim, "*that presence arises from the immersion of the system (the sensorimotor and effective valid actions it supports), the coherence*

**Figure 1: Relationships between presence concepts as proposed by Skarbez et al. (2017, 96:23); layout re-designed by the authors and framed as "*Skarbez-model*".**

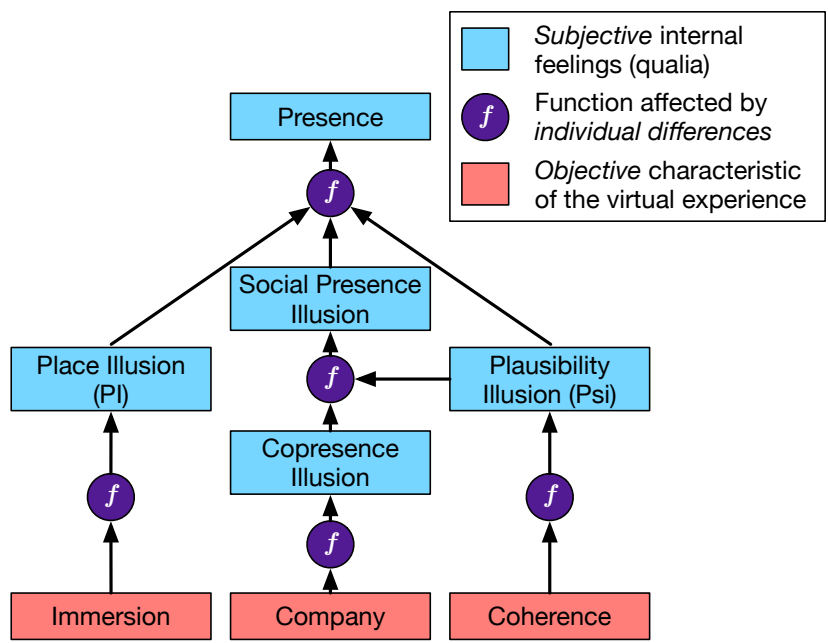

*of the scenario, whether the virtual experience offers company to the user, and the individual characteristics of the user.*" (96:23).

## Discussion of Current Presence-Oriented XR Theories

The proposed model by Skarbez (from now on Skarbez-model) is a well-motivated extension of the older two-component model by Slater (from now on Slater-model) based on the PI and the Psi, and immersion as the sole two objectively measurable (system) characteristic. Specifically, their introduction of coherence as a separate (measurable) characteristic opens up interesting perspectives. In addition, the identification of the influence of the Psi on social presence is well motivated. The Skarbez-model also integrates various findings from the literature about the many different aspects of presence, e.g., concerning social and co-presence, and hence fosters the understanding of some of the primary constructs relating to the study of virtual experiences. However, we argue that there are still potential theoretical and conceptual difficulties with the Skarbez-model, some rooting back to the older Slater-model as a precursor, e.g., when we make a distinct argument against the usage of the term illusion for qualia. We start the discussion by a set of questions about Skarbez-model's propositions:

### I. Questions about the selection of constructs and their relations

1. Why are qualia arranged hierarchically? Shall it imply the feeling of being with someone to be *less important* than the feeling of being there? Any importance does not emerge from a theoretical order but from the kind of interaction, the kind of experience per se. Other qualia, such as the virtual body ownership illusion (VBOI) seem excluded arbitrarily, despite its indicated impact on presence, see, e.g., (Waltemate et al., 2018).
2. Why is the Psi affecting presence and social presence but not the PI? We argue that a successful PI is affected by the coherence and plausibility of spatial cues. Hence, the Skarbez-model seems overly restrictive in its integration of plausibility in the overall theory, and, similarly, for its integration of coherence only contributing to the Psi.

### II. Questions about construct layers and construct status

3. Why is presence in the Skarbez-model defined based on the perceived realness of an XR experience? We agree that realness in the sense of "in coherence with sensory stimuli by natural sources" plays a critical role on the sensory layer to achieve sufficient ergonomic qualities, e.g., to avoid unwanted effects like cybersickness (Stauffert et al., 2018, 2020). However, on higher levels, e.g., the cognitive layer, presence can be evoked via the PI by simple line renderings not resembling any real objects by form, color, or detail.
4. Why are the specific presence-related constructs called illusions? A quale is by definition a subjective conscious experience. From a perceptual point of view, an illusion occurs when a subjective perception lacks an objective representation. But XR provides perceivable objective representations corresponding to subjective perceptions. In this sense, the Skarbez-model does identify presence as a quale and not an illusion but fails to do this for the contributing qualia, i.e., place illusion, plausibility illusion, social presence illusion, and copresence illusion.

Some of these criticisms go far beyond a mere terminological debate and cannot be counteracted by simple extension of the model. For example, when we talk about illusions throughout such models then **we are conceptually manifesting the overall separation into reality and virtuality as a form of deception**. However, our models should be capable of convincingly describing where we assume the transfer from artificially generated stimuli to qualia occurs, and that the effects on the users are indistinguishable from similar effects caused by natural (non-artificial) stimuli. That does not imply that people do not know that they are in an artificial environment (as in the film *The Matrix*).

Phenomenological, artificial objects and environments engender a proximate stimulus representation that corresponds to subjective perception. **Besides, any subjective perception and experience, any qualia, must be assumed as real.**

Skarbez et al. (2017, 2020) also reflected on said illusion problem. They defined a quale to focus on perceived realness in contrast to actual realness as a function of a system's ability to provide stimuli that match reality, i.e., a function of immersion and coherence. They also suggested discriminating between the Place Illusion as an illusory (false) feeling of being in a remote or virtual place and placeness as a feeling of being in a real place. We argue that there is nothing like a "*false*" feeling. A quale is a subjective internal feeling which cannot be false or not real, at least from a phenomenological point of view. The sensory stimulations giving rise to a quale can be artificial, but do not render the effect "*false*", nor do they make the artificial stimulation "*false*". It is pragmatically just a distinction between the processes that generated said stimuli. Given a sufficient coherence between the quality of an artificial stimulus and the required or expected qualities as defined by our sensory, perceptual, and cognitive information processing layers, this distinction can subjectively vanish.

Similarly, regarding the second question above, one can argue that the introduced objective characteristic of coherence affecting the Psi which then affects the social presence illusion and presence but not the PI is motivated by model-specific definitions of the concepts of coherence and plausibility. If one restricts the latter ones to only impact on a cognitive level, then it is easier to argue that they don't necessarily also affect the PI. This makes the proposed model valid internally. Nevertheless, the introduction of concepts and terms to explain empirical findings should be done with care. One can, of course, define specific meanings to chosen terms upfront to precisely describe the intended interpretation. However, specifically with terms that have a common and widely used meaning, we would argue that it is best to stick with these definitions to strive for easy cognitive accessibility and make a model as much self-descriptive as possible. In this sense, we feel the Skarbez-model's concepts of coherence and plausibility to be partly misleading. They seem not to capture all potential applications within a presence theory and to be restricted to a subset of concepts. For example, coherence of artificial visual stimuli with spatial cues expected on the sensory and perception layer can lead to a plausible evocation of spatial self-orientation and – depending on the degree of the substitution of visual stimuli from the physical environment around the user – to an evocation of the feeling of "*being there*". Here, "*there*" would refer to a cognitive attribution of the sum of all spatial stimuli as to belong to an environment different from the physical environment around the user.

We honestly value the models by Skarbez et al. (2017) and by Slater (2009) and any predecessors not discussed here. Our criticisms are meant to motivate discussions and advancements in the development of theoretical models of XR experiences. From an HCI view, such models should not only generate a consistent theory of the interrelation and potential influences of important constructs, factors, and characteristics. They should also support guidelines for designers and developers to exploit the vast design space of XR experiences and their impact on human behaviors. This includes predictable impact paths and systematically measurable and manipulable variables (Wienrich et al., 2021) to acquire knowledge with practical impact.

## Beyond Presence: Congruence and Plausibility (CaP)

This section proposes an alternative model of XR experiences. It builds upon Skarbez et al. (2017) and Slater (2009), taking the raised criticism into account. It also shifts away from presence, i.e., the sense of "*being there*" (the PI), as the central quale to capture the many variations of XR covered by the Virtuality Continuum (VC) (Milgram and Kishino, 1994). The concept of the PI gets more and more blurred once we move along the VC towards the non-simulated real environment. At which point do we know that we are dealing with a Place Illusion, i.e., something that is mainly caused by simulated content, and when do we have to accept that spatial cues making us feel to be in a place are not simulated but stem from the real environment? Hence, in the wider scope of Mixed and Augmented Reality (MR & AR), the PI becomes much less prominent.

Besides, XR technology and application development progress continuously, and its quality should likewise be evaluated. XR is already applied as a therapy system, mind, and behavior changer. Hence, we already know and accept that XR can bring users (real) experiences and causing (real) behavior. It might be comparable to the pragmatic quality in the user experience research. We presuppose that a technical device fulfills a specific function, but we are additionally interested in the hedonic, eudemonic, or social quality following the interaction with the device (Wienrich and Gramlich, 2020).

We follow Slater (2009) and Skarbez et al. (2017) and adopt *plausibility* as the first component. Valid alternatives for plausibility include *acceptance* or *suspension of disbelief* (Cruz-Neira et al., 1992; Heeter, 1992) but we focus for now on plausibility in analogy to former theoretical models. We also further specify coherence as *congruence* and include it as the second component of our proposed Congruence and Plausibility (CaP) model. Here, **congruence is describing the objective match between processed and expected information on the sensory, perceptual, and cognitive layers.**[1]

However, in contrast to the discussed presence models, we don't assume an illusion of plausibility **but define plausibility as a state or condition during an XR experience that subjectively results from the evaluation and congruence of information processed by the sensory, perceptual, and cognitive layers.** In our CaP-model, congruence and plausibility become central components affecting information processing on every level and giving rise to the acceptance and the suspension of disbelief (Heeter, 1992). Figure 2 illustrates the conceptual view of the proposed CaP-model including the main components and their relations.

---

[1] Earlier versions of the CaP-model were still relying on coherence as its second component and first published follow-up works have adopted this. This is still valid since we see congruence as an ontological specification of coherence.

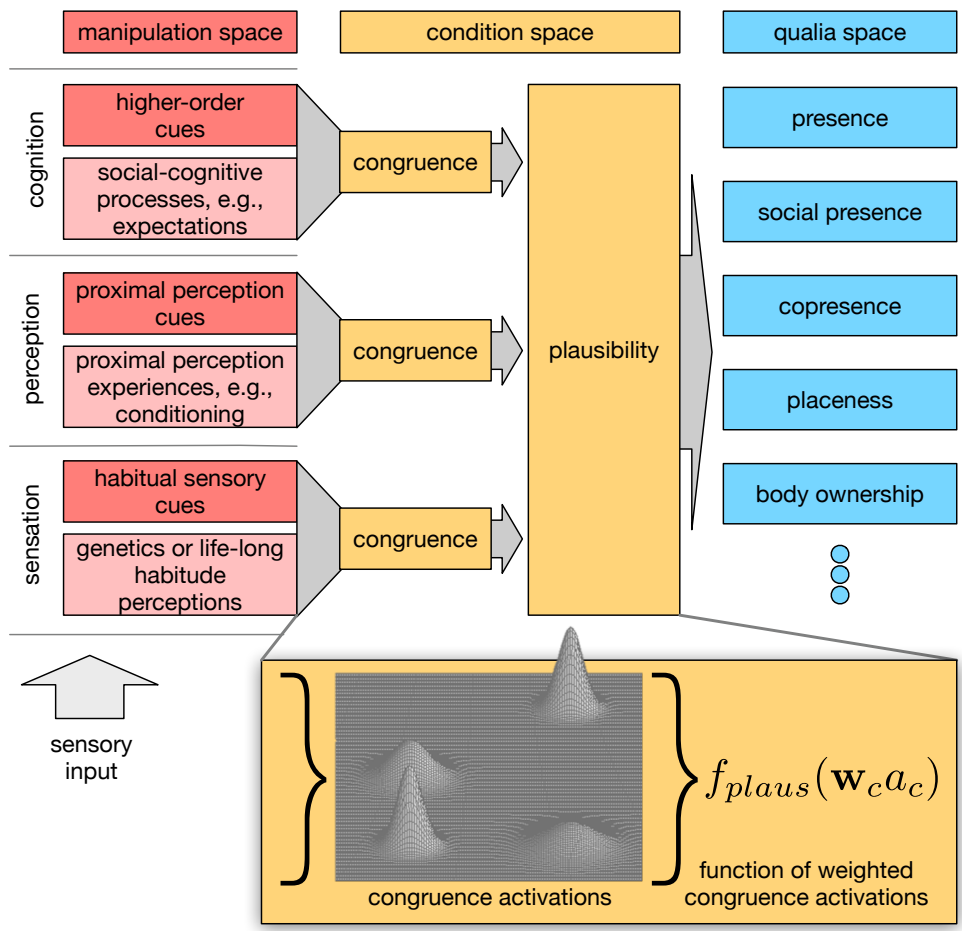

**Figure 2: Relationship between different XR-related qualia (including presence) and the contributing sensory information and cues proposed as an alternative new theoretical model for XR experiences and the related components. Congruence and plausibility of cues on the sensation, perception, and cognition levels take on a central role between the design and manipulation space of XR experiences and the evoked qualia. Plausibility emerges from a function of weighted congruence activations.**

The model assumes that plausibility arises from the congruence of cues on each of these layers. Each layer sets up *a frame* that defines the congruence conditions of how information is processed and interpreted and to which extent cues can be considered congruent. Here, the sensory layer exposes the base frame of information processing by setting the boundary conditions of how we transduce physical and physiological signals into neural signals. Permanently changing this frame is mainly restricted to genetic and epigenetic adaptions or cyber implants. Temporary modulation would include neuro-active drugs. The congruence conditions on this layer are accessible from biological and physiological knowledge.

In contrast to the sensory layer, the frames for the interpretation of sensory information on the perceptual and cognitive layer exhibit much more accessible plasticity and manipulation space since they are additionally also shaped by the recipient's learning, memory, knowledge, mental model, expectation, and attention, i.e., proximal perception experiences and social-cognitive processes. Imagine simple animated line drawings on a 2D display. If the resulting patterns match comparable patterns generated by a perspective projection of forward/backward movements in a 3D tunnel, the resulting perceptual congruency evokes vection independently of the underlying process generating the percepts, or any degree of realness or vividness. An example of cognitive congruency is a potential appearance match of a user's avatar with her/his real physical appearance. While there is evidence that an increased match increases factors of presence or emotional response (Waltemate et al., 2018), or acceptance (Latoschik et al., 2017), an absolute congruence is not necessary to accept the virtual body as one's own as demonstrated by the Proteus effect (Yee and Bailenson, 2007).

Congruency is constituted by relations between the cues and the XR experience itself. The experience can be congruent in relation to the habitual sensory cues, proximal perceptual cues, or higher-order cognitive cues. Plausibility emerges from a function of weighted congruence activations. A weighted process models dynamically changing contributions of congruent and/or incongruent relations. For example, the narrative or the use (cognitive layer) of an XR experience can be quite compelling. Then lower sensory congruencies might contribute less strongly to plausibility and the corresponding quale. Besides, at least the sensory level of maximal congruence is reached at a certain technical advancement, since users' given sensory capabilities can be considered fixed. Thus, with a certain level of technological development, the level of congruence stemming from sensory relations is constantly high, but the contribution to the plausibility emergence is still variable.

The distinction into the different sensory information processing layers allows to pinpoint how congruence affects evaluation given a respective frame. It provides a clearer picture of the interrelated components while it is in line with Slater's definitions of the PI to be constrained by the sensorimotor contingencies, i.e., how the world is perceived, and the Psi as the illusion that the scenario being depicted is actually occurring, i.e., what is perceived (Slater, 2009). The different cue levels, reaching from bottom-up to top-down, enable prediction and empirical testing of the resulting congruence and plausibility conditions. While the bottom-up framed congruence is primarily measured objectively and quantitatively, the top-down framed congruence is mostly assessable by subjective ratings, qualitative observations, or deceiving behavioral observations. However, the suggested XR experience model allows for systematic a priori predictions and post-hoc explanations.

The proposed model also does not need to further define the resulting qualia's exact meaning and is largely independent of this. In other words, the model is valid for those qualia researchers, designers, and developers are interested in. The procedure to predict a priori or explain post-hoc the relation between the manipulated cues and the conditions experienced in XR remains the same. For example, if a definition requires us to specify a certain degree of realism (as in the Skarbez-model) then it is up to the defining instance to specify the assumed layer(s) and respective cues precisely and designers can check if they can generate such cues in congruence with the expected qualities on that layer.

## Discussion

This article proposed the CaP-model of XR experiences based on congruence and plausibility as central components. The proposal derived the central ideas and concepts from an analysis of promising components and of potential shortcomings of existing models by Slater (2009) and later Skarbez et al. (2017). We conclude with an assessment of our model regarding important requirements (typeset in *italics*) of such a model before we discuss limitations.

In our opinion, the CaP-model possesses *predictive and explanatory power* of modern XR experiences. The manipulation space offers realizable and systematically controllable manipulations. Well-defined frames of interpretation of the cues enable congruence checks and then a priori predictions or post-hoc explanations of the influence of those cues on the plausibility condition and hence the corresponding qualia. For example, if the sensory layer determines how something is sensed, objective congruence tests can assure the desired quality (e.g., to assure a required frame rate or similar technical characteristics). On higher levels, user testing might be better suited. However, despite testing for the many potential qualia, we now only have to primarily test for plausibility of the cues defined upfront as being required to evoke a certain quale.

Further, our CaP-model integrated *the body of knowledge on presence and related XR constructs*. Simultaneously, it is able to *avoid the aforementioned potential shortcomings of the existing model(s)*. It arranges XR experience-related qualia at one level and postulates plausibility as one common constituting and pivotal factor and a corresponding testable approach of its emergence. It shifts the focus from place illusion and centered three cue layers influencing congruence and plausibility and then the considered qualia. Thus, the model proposes the same prediction paths, also resolving the question of inter-qualia-correlations. It resolves the often-inefficacious debate about the comparison with real-world experiences or the realness of XR experiences by accepting that XR is capable of bringing users (real) experiences and causing (real) behavior. In this sense, the proposed model identifies presence as a quale and not as an illusion and does this for other contributing qualia, such as social presence, copresence, placeness, and body ownership. Notably, we define plausibility as a *true* and for the user *real* condition during a XR experience rather than an illusion making the operationalization much easier. Questions can be formulated directly and do not rely on *as-if* comparisons.

Similarly, our CaP-model also incorporates the valid and necessary distinction between qualia and objectively measurable characteristics (Slater and Wilbur, 1997, 8), e.g., as intended by the identification and definition of factors like immersion (Slater, 1999) or company and coherence (Skarbez et al., 2017). However, our proposed model essentially simplifies such influences by identifying them as variations of just one factor our model integrates as congruence, but in a much broader context compared to (Skarbez et al., 2017) since the model incorporates congruence on all three layers of sensation, perception and cognition.

## Limitations

The present contribution is meant as a position paper taking empirical data *verifying or falsifying* the model out of scope. However, the present paper is a solid base for a set of such experiments in the future. Similarly, the *validity and soundness* requirements must be tested in future studies as well.

Finally, our proposed model simplifies complex processes, as each model that tries to predict and explain human experience will have to do to a certain extent. The proposed model purposely does not claim any further details about the dependencies or interrelations between the different qualia and the resulting structure, e.g., a hierarchy of factors contributing to the overall construct of presence as proposed by (Skarbez et al., 2017). As we noted, in a recent experiment, we manipulated presence and measured a correlating change in virtual body ownership, and vice versa, giving rise to speculation of an additional latent constituting factor affecting both. The latter approach highlights how these potential relationships can be investigated and it already hints to a more complex interplay of components where the functional dependency is a) not directed unilaterally and/or b) hints to additional latent factors yet to be found. However, at this stage our proposed model purposely does not try to further highlight any qualia interrelations (on the right side of Figure 2) since it focuses on hypothesized congruences evoking plausibility of surrounding space, embodiment, company, social interaction, and the like. Simplifications risk implicating imprecision and a lack of detail. However, they simultaneously are a necessary prerequisite for a successful generalization which in turn helps to facilitate understanding and practical usage.

## Conflict of Interest

The authors declare that the research was conducted in the absence of any commercial or financial relationships that could be construed as a potential conflict of interest.

## Author Contribution

The authors have contributed equally to this work.

## Funding

This publication was supported by the Open-Access Publication Fund of the University of Würzburg, by the XR Hub Bavaria, the German Ministry of Science and Education (BMBF) in the projects ViTraS (grant 16SV8219) and HiAvA (grant 16SV8781), and the EU in the project VIRTUALTIMES (grant 824128).